\documentclass[%
 reprint,
 amsmath,amssymb,
 aps,
]{revtex4-1}

\usepackage{graphicx}% Include figure files
\usepackage{dcolumn}% Align table columns on decimal point
\usepackage{bm}% bold math
\begin{document}

\preprint{APS/123-QED}

\title{Bi-hyperbolic isofrequency surface in a~magnetic-semiconductor superlattice}%

\author{Vladimir R. Tuz}
\affiliation{International Center of Future Science, State Key Laboratory on Integrated Optoelectronics, College of Electronic Science and Engineering, Jilin University, 2699  Qianjin Str., Changchun 130012, China
}
\affiliation{Institute of Radio Astronomy of National Academy of Sciences of Ukraine, 4, Mystetstv Str., Kharkiv 61002, Ukraine}
\author{Volodymyr I. Fesenko}%
 \email{volodymyr.i.fesenko@gmail.com}
\affiliation{International Center of Future Science, State Key Laboratory on Integrated Optoelectronics, College of Electronic Science and Engineering, Jilin University, 2699  Qianjin Str., Changchun 130012, China
}
\affiliation{Institute of Radio Astronomy of National Academy of Sciences of Ukraine, 4, Mystetstv Str., Kharkiv 61002, Ukraine}

\author{Illia V. Fedorin}
\affiliation{National Technical University `Kharkiv Polytechnical Institute', 2, Kyrpychova Str., Kharkiv 61002, Ukraine}%

\date{\today}% It is always \today, today,

\begin{abstract}
The topology of isofrequency surfaces of a magnetic-semiconductor superlattice influenced by an external static magnetic field is studied. In particular, in the given structure, topology transitions from standard closed forms of spheres and ellipsoids to open ones of Type I and Type II hyperboloids as well as bi-hyperboloids were revealed and analyzed. In the latter case, it is found out that a complex of an ellipsoid and bi-hyperboloid in isofrequency surfaces appears as a simultaneous effect of both the ratio between magnetic and semiconductor filling factors and magnetic field influence. It is proposed  to consider the obtained \textit{bi-hyperbolic} isofrequency surface as a new topology class of the wave dispersion.
\end{abstract}

\pacs{Valid PACS appear here}% PACS, the Physics and Astronomy
\maketitle

%\tableofcontents

A topology of photonic systems is related to global behaviors of a wave function accounting constitutive and structural parameters in the entire dispersion band \cite{Ling:14}. It means that in the phase space of a propagating electromagnetic wave at a constant frequency the topology appears in the form of an isofrequency surface (also known as Fresnel wave surface or surface of wave vectors), which governs the wave propagation conditions along an arbitrary direction inside the corresponding optical material (in fact it expresses the relationships between the directions of the wave vector and the vector of group or phase velocity of the wave; for the first reading on construction of isofrequency surfaces according to the laws of geometrical optics we refer to methodological notes in \cite{lock_PhysUsp_2008}). By definition, each isofrequency surface belongs to a class of \textit{quadrics} \cite{Coxeter_1963}, from a large variety of which three particular nondegenerated forms are well known in optics--sphere, ellipsoid and hyperboloid. 

Thus, in an isotropic medium isofrequency surfaces appear in the closed form of a sphere, whereas in a uniaxial optical crystal they transit to a complex of a sphere and spheroid, which characterize propagation conditions of ordinary and extraordinary waves, respectively \cite{Born_1968}. These isofrequency surfaces can intersect each other in some sections at certain singular points. In a biaxial crystal the complex consists of a sphere and ellipsoid. In other natural anisotropic media including acoustic crystals, plasmas and magnetically ordered (gyrotropic) media isofrequency surfaces can acquire both closed and open forms. In the latter case they resemble the form of a hyperboloid (see Fig.~8.3.2 in \cite{felsen1994radiation} for a taxonomy of isofrequency surfaces in anisotropic media). In this way, different forms of topology of the wave dispersion express the kind of anisotropy, namely the relations between components of permittivity and/or permeability tensors characterizing the medium. 

\begin{figure}
\centering
\includegraphics[width=\linewidth]{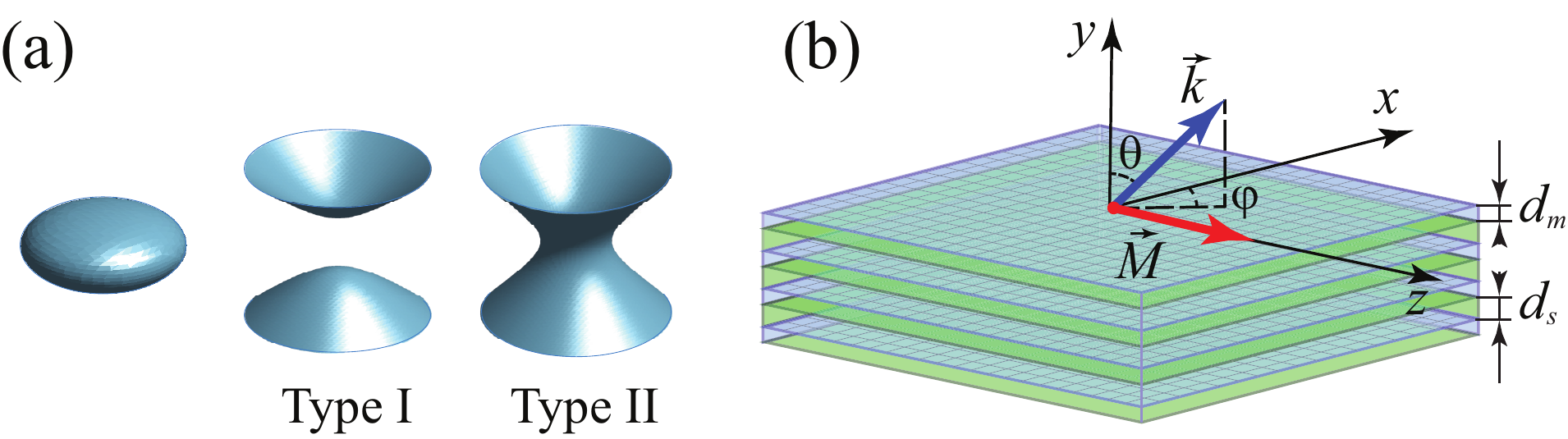}
\caption{(a) Three typical topologies (ellipsoid, two-fold Type I and one-fold Type II hyperboloids) of isofrequency surfaces inherent in anisotropic media, and (b) the problem sketch related to a magnetic-semiconductor superlattice influenced by an external static magnetic field.}
\label{fig:sketch}
\end{figure}

Indeed, in an anisotropic crystal when all principal values of its permittivity tensor are positive (i.e., $\varepsilon_{\parallel}>0$ and $\varepsilon_{\perp}>0$), the isofrequency surfaces have closed forms [Fig.~\ref{fig:sketch}(a)]. Contrariwise, when one or two corresponding tensor's components are negative (i.e., the medium is ``extremely'' anisotropic), the topology appears in the open form of two-fold (Type I with $\varepsilon_{\parallel}<0$ and $\varepsilon_{\perp}>0$) or one-fold (Type II with $\varepsilon_{\parallel}>0$ and $\varepsilon_{\perp}<0$) hyperboloid \cite{Poddubny:13}. Moreover, in a biaxial hyperbolic crystal, when all principal values of permittivity or permeability tensor are different and one of these values is negative, the isofrequency surface for the extraordinary wave possesses the form of an asymmetric hyperboloid.  

In natural media the hyperbolic topology can appear in gyroelectric (e.g., plasma) and gyromagnetic (e.g., ferrite) materials. By applying a strong external magnetic field to such materials, in a certain frequency band their corresponding dielectric or magnetic tensor becomes extremely anisotropic, and the propagation of electron or spin waves acquires hyperbolic isofrequency behaviors \cite{Fisher_PhysRevLett_1969, Lokk2017}. It results in the manifestation of specific features of wave propagation in such materials including the phenomena of nonreciprocity and unidirectional (one-way) topological transitions, as well as arising several reflected or refracted beams at the singular points of isofrequency surfaces, which are of great practical importance \cite{Berry200713, Hasan_RevModPhys_2010, Turpin:12, Ballantine_PhysRevA_2014, Leviyev_APLPhot_2017}. 

Despite the fact that hyperbolic topology can be found in natural anisotropic media, there is a considerable interest in creating artificial structures (metamaterials) possessing desired functionality. In practice, in order to construct a hyperbolic metamaterial, it is proposed to combine together metallic and dielectric counterparts into a unified structure. Two basic designs of such composites are widely discussed in the literature, namely a superlattice consisting of deep subwavelength alternating metallic and dielectric layers \cite{Wood_PhysRevB_2006, Zhukovsky:13}, and a lattice of metallic nanowires embedded into a dielectric host \cite{Belov_PhysRevB_2003}. In both configurations, the effective medium limit is usually implied. It is revealed that in such extremely anisotropic structured materials, thanks to the hyperbolic topology, strong enhancement of spontaneous emission, diverging density of states, negative refraction and enhanced superlensing effects can be reached \cite{jacob_ApplPhysLett_2008, Takayama_OptLett_2012}.

On the other hand, the fact that hyperbolicity requires plasma behavior in a certain direction of wavevector space and insulating behavior in the others leads to an option instead of metals along with dielectrics use some combination of semiconductors and magnetic materials (e.g., ferrites) as building blocks of the hyperbolic metamaterials \cite{Kruk:2016}. This possibility becomes even more attractive considering that influence of an external magnetic field to such magneto-active structures allows to gain a control over waves dispersion features \cite{Fesenko_OptLett_2016, Tuz_Superlattice_2017, Tuz_JApplPhys_2017}, nonreciprocal effect \cite{Camley_SurfSciRep_1987, temnov2010active}, and on their topological transitions \cite{Li_ApplPhysLett_2012, Chern:17}. Such systems are important for a number of practical applications in integrated photonic devices for telecommunications (see, for instance, \cite{Armelles_AdvOpticalMater_2013} and references therein).

Therefore, it is expected that combining together magnetic and semiconductor materials within a unified structure can results in arising new forms of topology of isofrequency surfaces. In particular, it is our goal to demonstrate in the present Letter, for the first time to the best of our knowledge, that depending on the ratio between filling factors of magnetic and semiconductor subsystems within a superlattice and the direction of an external static magnetic field with respect to the structure periodicity, a complex of isofrequency surfaces in the form of ellipsoid and bi-hyperboloid can be obtained. We consider the bi-hyperboloid isofrequency surface as a new class of topology of the wave dispersion that is unachievable in ordinary anisotropic media.

Thereby, further we are interested in isofrequency surfaces of a multilayered structure (superlattice) whose periodicity is infinitely extended along the $y$-axis [Fig.~\ref{fig:sketch}(b)]. It consists of alternating magnetic (with constitutive parameters $\hat \mu_m$, $\varepsilon_m$) and semiconductor (with constitutive parameters $\mu_s$, $\hat \varepsilon_s$) layers having thicknesses $d_m$ and $d_s$, respectively, whereas the structure period is $d = d_m + d_s$. It is assumed that the magnetic and semiconductor layers are magnetized uniformly by an external static magnetic field $\vec M$ which is aligned along the $z$-axis transversely to the structure periodicity.

It is supposed that all characteristic dimensions, namely the thicknesses of the magnetic and semiconductor layers as well as the period of the given structure satisfy the long-wavelength limit, i.e. they are all much smaller than the wavelength in the corresponding part of the superlattice ($d_m\ll \lambda$, $d_s \ll \lambda$, $d \ll \lambda$), and, thus, the multilayered structure is considered to be a finely-stratified one. In view of this assumption, the standard homogenization procedure from the effective medium theory \cite{Agranovich_SolidStateCommun_1991} is applied in order to derive averaged expressions for effective constitutive parameters of the multilayered system. Taking into account the precautions for the accuracy of the approximation \cite{Orlov_PhysRevB_2011, Kidwai_PhysRevA_2012}, the results of homogenization procedure are verified by using the rigorous transfer matrix technique \cite{Tuz_JOpt_2015}, and a good accordance is found for the chosen structure parameters and frequency.

Therefore, the whole structure is approximated by an infinite (bi)gyrotropic space, which is then characterized by relative effective permeability $\hat\mu_{eff}$ and relative effective permittivity $\hat\varepsilon_{eff}$ represented as the second-rank tensors $\hat\mu_{eff}=\left[ \mu_{xx},~\mu_{xy},~0;~-\mu_{xy},~\mu_{yy},~0;~0,~0,~\mu_{zz}\right]$ and $\hat \varepsilon_{eff}=\left[ \varepsilon_{xx},~\varepsilon_{xy},~0;~-\varepsilon_{xy},~\varepsilon_{yy},~0;~0,~0,~\varepsilon_{zz} \right]$. All expressions of these tensors components derived via underlying constitutive parameters of magnetic ($\hat \mu_m$, $\varepsilon_m$) and semiconductor ($\mu_s$, $\hat \varepsilon_s$) layers one can find in \cite{Tuz_JMMM_2016}. In particular, for our further calculations we follow the results of \cite{Wu_JPhysCondensMatter_2007}, where a magnetic-semiconductor multilayered structure is made in the form of a barium-cobalt/doped-silicon superlattice for operating in the microwave part of spectrum. Remarkably, the characteristic resonant frequencies of the underlying magnetic and semiconductor materials of the given superlattice are different but, nevertheless, closely spaced within the same frequency band. Thus, the resulting homogenized medium appears as an unbounded biaxial \textit{gyroelectromagnetic} crystal in which one of the optical axes is directed along the structure periodicity (the $y$-axis), while the second one coincides with the direction of the external magnetic field (the $z$-axis).

Let us consider a plane uniform electromagnetic wave with frequency $\omega$ and wave vector $\vec{k}$ which propagates in such a gyroelectromagnetic medium along an arbitrary direction. Time and space harmonic variations of the electric ($\vec e$) and magnetic ($\vec h$) fields are then given by
\begin{equation}
\vec e (\vec h) = \vec e_0 (\vec h_0) \exp\left[i \left(-\omega t + k_x x + k_y y + k_z z\right)\right],
\label{eq:space_time}
\end{equation}
where $k_x=k \sin\theta \cos\varphi$, $k_y=k \sin\theta \sin\varphi$, and $k_z=k \cos\theta$ are projections of the wave vector $\vec{k}$ in the Cartesian coordinates.

From Maxwell's equations written with respect to the field components one can derive a system of two equations that contains only $z$-components of the field \cite{Gurevich_book_1963}. Then, from this system the bi-quadratic dispersion equation describing propagation of electromagnetic waves through an unbounded gyroelectromagnetic medium can be obtained (the corresponding treatment for a biaxial crystal one can find in Chapter~11 of \cite{landau_1960_8}). In terms of $\kappa = k/k_0$ it has a form
\begin{equation}
A\kappa^4 + B\kappa^2+C=0,
\label{eq:disp_eq}
\end{equation}
whose solution is straightforward: $\kappa^2=(-B\pm\sqrt{B^2-4AC})/2A$. Here $A=(\varepsilon_{zz}\mu_{zz})^{-1} \left(\overline{\varepsilon} \sin^2\theta + \varepsilon_{zz} \cos^2\theta \right) \left(\overline{\mu} \sin^2\theta + \mu_{zz} \cos^2\theta\right)$, $B=-(\varepsilon_{xx}\mu_{yy} + \mu_{xx}\varepsilon_{yy} - 2\varepsilon_{xy}\mu_{xy})\cos^2\theta +(\varepsilon_{zz}\mu_{zz})^{-1}(\varepsilon_{\bot}\overline{\mu}\mu_{zz}+\mu_{\bot}\overline{\varepsilon}\varepsilon_{zz})\sin^2\theta$, $C=\varepsilon_{\bot}\mu_{\bot}$, and $\overline{\varepsilon} = \varepsilon_{xx}\cos^2\varphi + \varepsilon_{yy}\sin^2\varphi$, $\varepsilon_{\bot} = \varepsilon_{xx}\varepsilon_{yy} + \varepsilon_{xy}^2$, $\overline{\mu} = \mu_{xx}\cos^2\varphi + \mu_{yy}\sin^2\varphi$, and $\mu_{\bot} = \mu_{xx}\mu_{yy} + \mu_{xy}^2$ are introduced as generalized effective constitutive parameters.

The isofrequency surfaces are related only to the real $\kappa$ (while considering the \textit{lossless} medium), i.e., generally, the dispersion features of the given structure are characterized by two isofrequency surfaces [i.e., two isofrequency surfaces exist under the condition that Eq.~(\ref{eq:disp_eq}) yields two real values $\kappa_1$ and $\kappa_2$ at the given frequency; a single isofrequency surface exists if only $\kappa_1$ or $\kappa_2$ is a real quantity while another wavenumber is an imaginary quantity]. In accordance with accepted notations \cite{felsen1994radiation}, the roots of (\ref{eq:disp_eq}) with upper sign ``$+$'' and lower sign ``$-$'' represent dispersion of the ordinary and extraordinary waves, respectively 

It is evident that the externally applied magnetic field can essentially change the dispersion characteristics of the structure under study since it simultaneously influences both permeability and permittivity of the medium. Appearance of the isofrequency surfaces depends also on the direction of structure periodicity and ratio between filling factors of the magnetic and semiconductor subsystems within the superlattice (we introduce corresponding filling factors as dimensionless parameters written in the form $\delta_m=d_m/d$, $\delta_s=d_s/d$, and thus $\delta_m+\delta_s = 1$). Nevertheless, the forms of isofrequency surfaces of interest can be basically explained considering only the principal values of tensors characterizing effective permeability $\hat\mu_{eff}$ and effective permittivity $\hat\varepsilon_{eff}$. 

\begin{figure}[ht]
\centering
\fbox{\includegraphics[width=\linewidth]{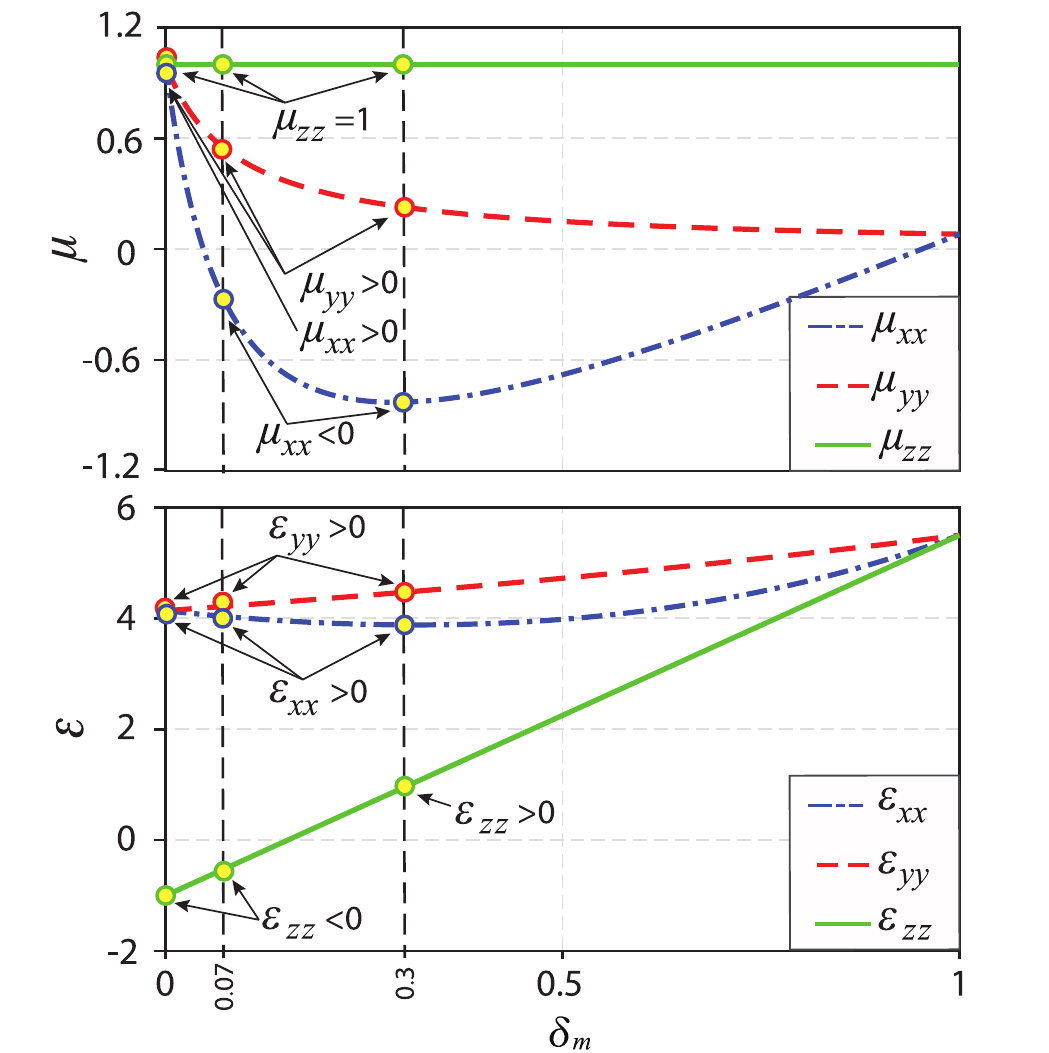}}
\caption{Principal values of the tensors of effective permeability  $\hat\mu_{eff}$ and effective permittivity $\hat\varepsilon_{eff}$ versus filling factor $\delta_m$. Following \cite{Wu_JPhysCondensMatter_2007} the magnetic-semiconductor structure under study is considered to be made from a set of barium-cobalt/doped-silicon layers designed for operation in the microwave part of spectrum. For the magnetic constitutive layers, under saturation magnetization of 2930~G, parameters are: $f_0=\omega_0/2\pi=4.2$~GHz, $f_m=\omega_m/2\pi=8.2$~GHz, $\varepsilon_m=5.5$. For the semiconductor constitutive layers, parameters are: $f_p=\omega_p/2\pi=10.5$~GHz, $f_c=\omega_c/2\pi=9.5$~GHz, $\varepsilon_l=1.0$, $\mu_s=1.0$. The frequency parameters is $k_0 = 155.5$~m$^{-1}$, that corresponds to $d/\lambda \approx 3 \times 10^{-2}$.}
\label{fig:principal}
\end{figure}

For further reference the principal values of both tensors are calculated at fixed frequency and plotted in Fig.~\ref{fig:principal} as functions of the filling factor $\delta_m$. For the given problem geometry, three representative combinations of the filling factor $\delta_m$ and corresponding principal values of constitutive tensors are distinguished considering where they acquire negative quantities, namely: (i) $\delta_m=0$, $\mu_{xx}>0$ and $\varepsilon_{zz}<0$; (ii) $\delta_m=0.3$, $\mu_{xx}<0$ and $\varepsilon_{zz}>0$; (iii) $\delta_m=0.07$, $\mu_{xx}<0$ and $\varepsilon_{zz}<0$. Note, two other principal values $\mu_{yy}$ and $\varepsilon_{yy}$ are positive quantities within the whole range of $\delta_m$, and $\mu_{zz}$ is a positive constant ($\mu_{zz}=1$). In Fig.~\ref{fig:principal} these combinations of interest are denoted by red, blue and green circles pinned along vertical dashed lines, which trace corresponding values of the filling factor $\delta_m$.   
 
\begin{figure}[!t]
\centering
\fbox{\includegraphics[width=\linewidth]{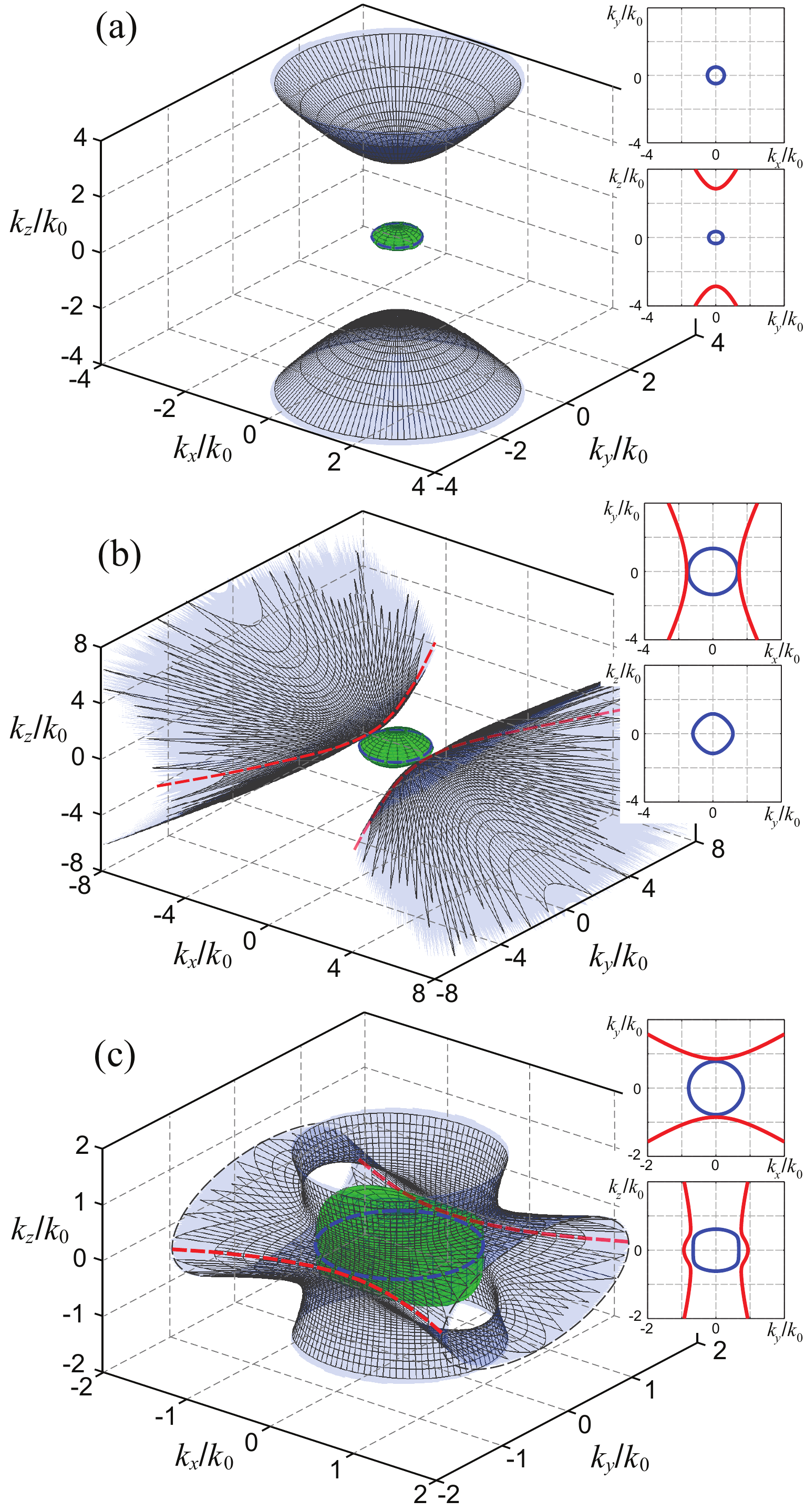}}
\caption{Transition of forms of isofrequency surfaces of a gyroelectromagnetic medium for different ratio between filling factors of magnetic and semiconductor subsystems; (a) $\delta_m =0.0$, $\delta_s=1.0$; (b) $\delta_m =0.3$, $\delta_s=0.7$; (c) $\delta_m =0.07$, $\delta_s=0.93$. All other parameters are as in Fig.~\ref{fig:principal}.}
\label{fig:isosurfaces}
\end{figure}

The isofrequency surfaces existing for three above chosen combinations of structure parameters are depicted in Fig.~\ref{fig:isosurfaces}. In this figure the isofrequency surfaces related to the ordinary wave are presented in green, while those related to the extraordinary wave are presented in gray. In the insets disposed in the right side of each plot the cross sections (contours) of corresponding isofrequency surfaces at $k_z=0$ and $k_x=0$ are additionally drawn. From this figure one can conclude that for three particular combinations of the filling factors $\delta_m$ and $\delta_s$ (when the frequency, external magnetic field strength, period, and other constitutive parameters of the underlying materials of the superlattice are fixed) the dispersion characteristics of the structure under study exhibit a remarkable difference in topology of the isofrequency surface related to the extraordinary wave, while that of the ordinary wave remains to have a typical closed form of a spheroid whose isocontour in the $k_x-k_y$ plane is a circle. 

In particular, for the first set of parameters, when $\delta_m = 0$, the structure represents a bulk semiconductor whose dispersion characteristics demonstrate a complex of isofrequency surfaces existing in the form of a spheroid for the ordinary wave and two-fold Type I hyperboloid for the extraordinary wave [Fig.~\ref{fig:isosurfaces}(a)]. Since in this case $\varepsilon_{zz}$ is a negative quantity, $\varepsilon_{xx}=\varepsilon_{yy}$, and $\mu_{eff}$ is a scalar ($\mu_{eff}=1$), both isofrequency surfaces are symmetric about the revolution axis oriented along the $z$-axis, which coincides with the direction of the external magnetic field.

For the second set of parameters, among all principal values of constitutive tensors only $\mu_{xx}$ is a negative quantity while other values are positive ones. In this case, the isofrequency surface of the extraordinary wave transits to the form of a two-fold Type I hyperboloid oriented along the extension of magnetic layers of the superlattice (i.e., along the $x$-axis in the given structure geometry) [Fig.~\ref{fig:isosurfaces}(b)] as is typical for the hyperbolic metamaterials \cite{Poddubny:13}. Nevertheless, there is a distinctive feature consisting in the fact that the hyperboloid appears to be slightly compressed along the $z$-axis, since in this direction an additional anisotropy axis arises as a result of the influence of the external magnetic field. We should note that at other frequencies, the isofrequency surfaces calculated for the first and second sets of parameters can appear also in the form of a one-fold Type II hyperboloid (not presented here).

Finally, for the third set of parameters, both $\varepsilon_{zz}$ and $\mu_{xx}$ are negative quantities, and the isofrequency surface related to the extraordinary wave transits to a combination of two one-fold Type II hyperboloids whose revolution axes are orthogonal, whereas the isofrequency surface of the ordinary wave remains to be in the form of a spheroid which lies inside of such a complicated hyperbolic form [Fig.~\ref{fig:isosurfaces}(c)]. In fact, this complicated form of the isofrequency surface of the extraordinary wave can be clearly explained by the effects of both external magnetic field influence (which is directed along the $z$-axis) and structure geometry related to the extension of magnetic layers of the given superlattice (which is directed along the $x$-axis). We consider that this obtained isofrequency surface can be related to the \textit{bi-hyperbolic} topology \cite{Ratzel_PhysRevD_2011}. It determines the wave dispersion features that can be found in gyrotropic media of a general class.

To conclude, we have examined the topology transitions of isofrequency surfaces existing in a magnetic-semiconductor superlattice which is influenced by an external static magnetic field. It is demonstrated, that in the given structure, a diversity of topologies (such as ellipsoid, two-fold Type I and one-fold Type II hyperboloids) can be achieved by changing the ratio between filling factors of magnetic and semiconductor subsystems within the superlattice providing the period and other constitutive parameters of the underlying materials are fixed. Moreover, we revealed that a complex of an ellipsoid and bi-hyperboloid in isofrequency surfaces appears as a simultaneous effect of both structure periodicity and magnetic field influence that can be considered as a new class of topological transitions.

\bibliography{state_isofrequency}

\end{document}